# Interpreting AFB$_b$, ALR and M$_W$

F. Richard

IJCLab

*Université Paris-Saclay, CNRS/IN2P3, 91405 Orsay, France*

______________________________________________________________________

**Contribution to the ECFA WG1-SRCH on Higgs/EW/top factories**

**Abstract**: *After the long-standing mismatch between ALR and AFBb measurements for predicting the Higgs boson mass, the recent and very precise result on Mw confirms the existence of a crisis in the sector of electroweak measurements. In this note, it is shown that ALR and Mw can be simply reconciled by assuming a common value for the T parameter. Consequently, the AFBb deviation is amplified and reaches a 4 s.d. significance. Interpreting this as evidence for a non-universal behaviour of the b quark is proposed within an extra dimension model. Large deviations with respect to the SM are expected at future Z factories.*

## Introduction

So far, the general moto has been: "**the SM works beautifully**, please circulate there is nothing to be noticed". At the time of LEP/SLD some tension had been noticed with the most precise measurements of $\sin^2\theta_W$ originating from two different observables: AFBb for LEP and ALR for SLD. While ALR only depends on Ae, AFBb goes like AeAb and could be affected by anomalies in the b sector.

The predicted mass for the lightest Higgs obtained by averaging of the two measurements, mH~90 GeV, came out in agreement with LHC, hiding a significant disagreement: while LEP result based on AFBb would predict a mass ~500 GeV, ALR from SLD predicted a mass around 30 GeV, with a mutual disagreement between the two $\sin^2\theta_W$ measurements at the 3.5 sd level. The recent result from CDF predicts Mh~20 GeV.

Taken at face value, the result from CDF ruins the alleged statement that SM is triumphant, and one can only dispute the experimental validity of this result.

The following figure, from [1], well displays these contradictions using the Higgs mass as a tool of comparison. One can see that almost all measurements, except AFBb, pull the result to a low mass, with an emphasis on ALR from SLD and Mw from CDF.

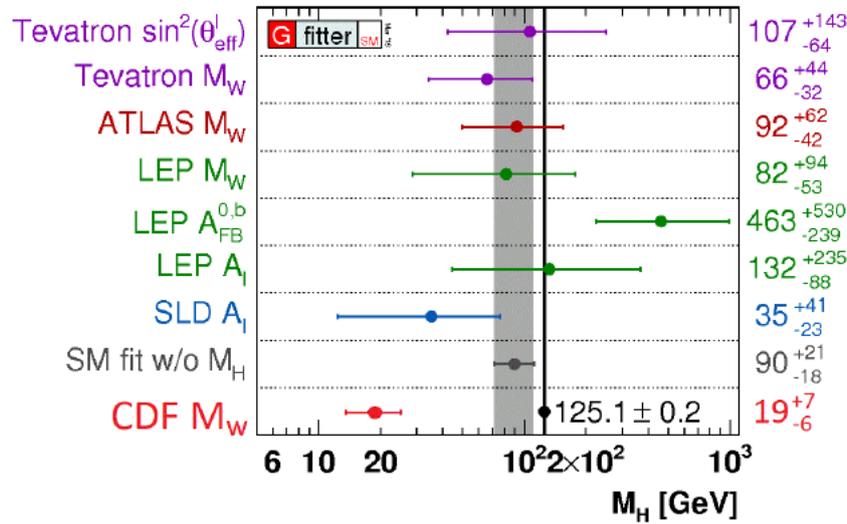

*Figure 1: Higgs mass predictions given by the most precise EW measurements performed at LEP, SLD, LHC and Tevatron.*

In the present note, I will assume the validity of these three results, ALR, AFBb, and Mw and show how ALR and Mw measurements can be reconciled with a common assumption on the Peskin Takeuchi parameter T. The consequence for AFBb is to deepen its excursion from the SM prediction, reaching a 4 s.d. significance. This suggests a non-universal behaviour of the b quark, which naturally occurs in composite and extra-dimensions models, as will be illustrated on an example.

## The T parameter interpretation for ALR and Mw

The remedy to the present situation is quite simple. Recall that BSM models can easily generate a sizeable T parameter and that one has:

$$\sin^2\theta_{\text{eff}} = \sin^2\theta_{\text{eff}}\big|_{\text{SM}} - \alpha \frac{c_w^2 s_w^2}{c_w^2 - s_w^2} T$$

Reconciling the SLD result with Mh=125 GeV is easily achieved by assuming that T=0.22±0.10, a large value but with a large uncertainty.

The CDF result on Mw also calls for the same large value of T, but with a much smaller error.

Recalling that:

$$M_W^2 = M_W^2\big|_{\text{SM}}\left(1 + \frac{s_w^2}{c_w^2 - s_w^2}\Delta r'\right),$$

$$\Delta r' = \frac{\alpha}{s_w^2}\left(-\frac{1}{2}S + c_w^2 T + \frac{c_w^2 - s_w^2}{4s_w^2}U\right)$$

one finds that **T=0.17±0.02** allows to reconcile Mw with the SM. It is therefore natural to assume this value for T, which corresponds to the following value of $\sin^2\theta_W$:

$$\sin^2\theta_W = 0.2311 \pm 0.0003$$

a result in perfect agreement with the Higgs mass value. This choice based on Mw from CDF, brings the ALR result to an agreement with SM, while the other measurements, less precise, remain consistent with the SM with the notable exception of AFBb.

## The origin of T

The origin of this large T value is of course an issue of major interest for predicting BSM physics. There are various possible answers to this question, and I will only give two examples.

In [2], within the N2HDM model, and assuming the light scalar h(95) indicated by LEP and CMS, the authors were able to reconcile the SLD and CDF measurements by generating the adequate T value.

During a presentation of these results at the ECFA workshop[1], one of these authors concluded that "*if the N2HDM is realized in nature and describes Mw from CDF, it is only in agreement with ALR, but not with AFBb, which therefore might be flawed.*"

The Mw result has generated a massive number of interpretations, which comfort the idea that it is trivial to generate appropriate corrections through T. Among these proposals stands the Georgi Machacek model, which is also invoked to explain indications for light scalars at LHC [3]. This model takes a clever combination of two triplets with the same vacuum expectation values to cope with the well-known $\rho \sim 1$ constraint, which is nothing else than assuming that T=0 at the tree level. It is intuitive to assume that this cancellation could be imperfect and generate the effect observed at CDF.

These examples are of course not exhausting a large domain of possibilities offered to phenomenology.

## Consequences on the AFBb result and possible interpretations

It is tempting to take at face value the SLD+CDF result and to pursue the reasoning by investigating the consequences of this result on the AFBb measurement.

Again, the sceptics will also discard this result with the usual sentences about the challenging aspects of such a measurement.

Nevertheless, let us proceed assuming that the detailed work of the 4 LEP experiments, under the supervision of a Nobel prize, can be trusted and assume the SLD+CDF result for $\sin^2\theta_W$.

An obvious question to ask is: what becomes of the disagreement between the SM prediction for AFBb and the experimental result from LEP?

---

[1] *1st ECFA Workshop* on Higgs/EW/Top factory, Oct 2022.

Not unexpectedly, the answer is that this disagreement, which was below the 3 s.d. level, jumps at 4 s.d. and therefore also becomes a serious contender of the SM.

Here I will give an example of a model which allows to explain such a large discrepancy. It interprets the flavour sector by assuming that the various flavours occupy different sites in an extra dimension, as displayed by the following picture:

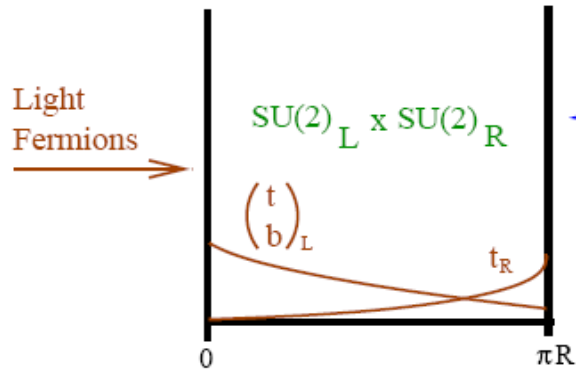

In this scheme, the doublet (t,b)L peaks near a brane, called the Higgs brane, and tends to behave as SM particles, while tR and bR tend to belong to the so called Planck brane, with couplings at variance with the SM. A detailed phenomenology can be found in [4] where a quantitative interpretation of the disagreement on AFBb was provided.

Given that AFBb=3/4AeAb, with AFBbSM=0.1067+-0.0008 and AFBb=0.00992±0.0016 from the LEP measurement, one deduces that:

$$dAb/Ab=-7\pm1.6\%$$

using an Ae deduced from SLD+CDF.

Recall that with ILC and polarized beams this quantity can be measured 50 times more precisely than at LEP [5].

Rb will also be measured, allowing to separate the L and R components as suggested by above picture.

## Measurements at future e+e- colliders

Future e+e- colliders should allow to improve dramatically the accuracies reached at the Z pole. This will also be the case in the continuum, meaning that LEP1 measurements need to be redone to match the predicted accuracies of the continuum. This is shown in [5], which also compares ILC and FCCee performances at the Z pole, as displayed in figure 2. Contrary to a common prejudice, this figure shows that beam polarisation allows to reach various observables inaccessible otherwise.

This feature offers the unique possibility to directly measure Ab from the combination:

$$ALRFB = \frac{1}{Pe}\frac{(\sigma F - \sigma B)L - (\sigma F - \sigma B)R}{(\sigma F + \sigma B)L + (\sigma F + \sigma B)R} = \frac{3}{4}Af$$

allowing a precise test of universality.

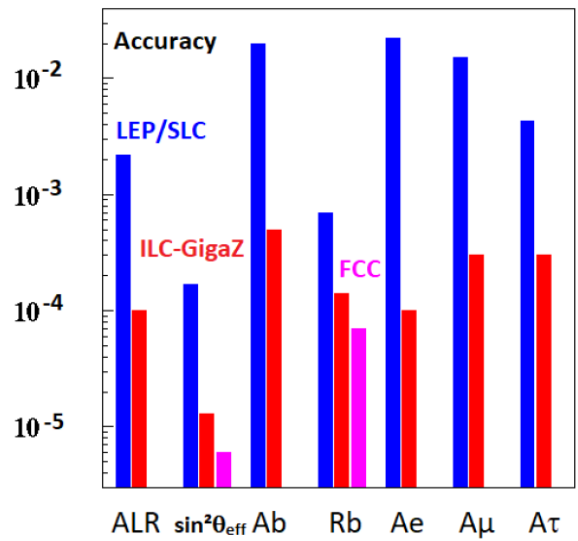

Figure 2: Estimated measurement errors at ILC-GigaZ compared to LEP1/SLC and FCCee.

More than an order of magnitude improvement with respect to LEP/SLD can be reached which will provide the final answer on the present interrogations.

Recall also that ILC promises a measurement of Mw, by threshold scan, with an error of ~2 MeV.

## Conclusion

In this short note, I have attempted to show how one can **reconcile the 3 most precise measurements of the electroweak parameters** and briefly sketched the consequences of these findings for future measurements which will be performed at e+e- machines. The main message seems to be that these measurements will give the final proof of the **presence of BSM physics** through the **T parameter** measurement and through a significant deviation of the **Ab** asymmetry from the SM. These effects need not be related but offer promising discoveries for the future of HEP.

**Acknowledgement:** Useful and lively discussions with Sven Heinemeyer, on various occasions, are gratefully acknowledged.